\documentclass[12pt]{article}
\newcommand{\conf}[1]{}

\newcommand{\squeezelist}{\setlength{\itemsep}{0pt}}
\def\a{{\alpha}}
\def\b{{\beta}}
\def\s{{\sigma}}
\def\t{{\theta}}
\def\e{{\epsilon}}
\def\d{{\delta}}

\def\v1{{v_{i+1}}}
\def\vi2{{v_{i+2}}}
\def\vn1{{v_{n-1}}}
\def\vp1{{v'_{i+1}}}
\def\pie{{\Pi_\epsilon}}
\def\pixy{{\Pi_{xy}}}
\def\P{{\cal P}}

\input{psfig} 
\newenvironment{pf}{\unskip{\bf Proof:}}{\unskip{\hfill $\Box$}}

\newcommand{\lemlab}[1]{\label{lemma:#1}}
\newcommand{\theolab}[1]{\label{theo:#1}}

\newcommand{\figlab}[1]{\label{fig:#1}}
\newcommand{\seclab}[1]{\label{section:#1}}

\newcommand{\lemref}[1]{\ref{lemma:#1}}

\newcommand{\theoref}[1]{\ref{theo:#1}}

\newcommand{\figref}[1]{\ref{fig:#1}}
\newcommand{\eqref}[1]{(\ref{eq:#1})}
\newcommand{\secref}[1]{\ref{section:#1}}

\newtheorem{theorem}{Theorem}[section]
\newtheorem{lemma}[theorem]{Lemma}

{\catcode`\@=11
\gdef\setft#1#2#3{%
\def\@oddfoot{
{\setbox0=\hbox{#1}
\setbox1=\hbox{#3}
\ifdim\wd0>\wd1
\dimen0=\wd0
\box0\hfil#2\hfil\hbox to\dimen0{\hfil\hfil\box1}
\else \dimen0=\wd1
\hbox to\dimen0{\box0\hfil }\hfil#2\hfil\box1 \fi
}}} }

\def\complaint#1{}
\def\withcomplaints{
\newcounter{mycomplaints}
\def\complaint##1{\refstepcounter{mycomplaints}%
\ifhmode%
\unskip%
{\dimen1=\baselineskip \divide\dimen1 by 2 %
\raise\dimen1\llap{\tiny -\themycomplaints-}}\fi%
\marginpar{\tiny [\themycomplaints]: ##1}}%
}

\usepackage{amssymb}
\usepackage{latexsym} 

\title{
{\bf Locked and Unlocked\\
Polygonal Chains in 3D}\thanks{This research was initiated at
	a workshop
	at the Bellairs Res. Inst. of McGill Univ.,
	Jan. 31--Feb. 6, 1998.
	This is a revised and expanded version
	of~\protect\cite{bddlloorstw-lupc3d-99}.
	Reseach supported in part by FCAR, NSERC, and NSF.
}
}
\author{
T.~ Biedl\thanks{
        University of Waterloo, Waterloo, Canada.
        \{biedl, eddemaine, mldemaine, alubiw\}@uwaterloo.ca.
}
\hspace{4mm}
E.~ Demaine\footnotemark[2]
\hspace{4mm}
M.~ Demaine\footnotemark[2]
\hspace{4mm}
S.~ Lazard\thanks{
	INRIA Lorraine, France.
	lazard@loria.fr.
}
\\
A.~ Lubiw\footnotemark[2]
\hspace{4mm}
J.~ O'Rourke\thanks{
        Smith College, Northampton, USA.
        \{orourke, streinu\}@cs.smith.edu.
}
\hspace{4mm}
M.~ Overmars\thanks{
	Utrecht University, The Netherlands.
	markov@cs.ruu.nl.
}
\hspace{4mm}
S.~ Robbins\thanks{
        McGill University, Montreal, Canada.
        \{stever, godfried, sue\}@cs.mcgill.ca.
}
\\
I.~ Streinu\footnotemark[4]
\hspace{4mm}
G.~ Toussaint\footnotemark[6]
\hspace{4mm}
S.~ Whitesides\footnotemark[6]
} 

\begin{document}
\maketitle

\begin{abstract}
\small
In this paper, we study movements of simple polygonal chains
in 3D.  We say that an open, simple polygonal chain can be
{\em straightened\/} if it can be continuously
reconfigured to a straight sequence of segments
in such a manner that both the length of each link
and the simplicity of the chain are maintained throughout
the movement.
The analogous concept for closed chains is {\em convexification\/}:
reconfiguration to a planar convex polygon.
Chains that cannot be straightened or convexified are called {\em locked}.
While there are open chains in 3D that are locked, we show that
if an open chain has a simple orthogonal projection
onto some plane, it can be straightened.
For closed chains, we show that there are unknotted but locked closed
chains, and we provide an algorithm for convexifying a planar
simple polygon in 3D.  All our algorithms require only
$O(n)$ basic ``moves'' and run in linear time.

\normalsize
\end{abstract}

\section{Introduction}
\seclab{Introduction}
A {\em polygonal chain\/} $P=(v_0,v_1,\ldots,v_{n-1})$ is a sequence
of consecutively joined segments (or edges) 
$e_i =v_iv_{i+1}$ of fixed lengths $\ell_i = |e_i|$, 
embedded in space.\footnote{
	All index arithmetic throughout the paper is mod $n$. 
}
A chain is {\em closed\/} if 
the line segments are joined in cyclic fashion, i.e., if $v_{n-1}=v_0$;  
otherwise, it is {\em open}.
A closed chain is also called a {\em polygon}.
If the line segments are regarded as obstacles, then the chains must remain 
{\em simple\/} at all times, i.e., self intersection is not allowed.
The edges of a simple chain are
pairwise disjoint except for adjacent edges, which share the common
endpoint between them.
We will often use {\em chain\/} to abbreviate ``simple polygonal chain.''
For an open chain our goal is to straighten it;
for a closed chain the goal is to {\em convexify\/} it,
i.e., to reconfigure it to a planar convex polygon.
Both goals are to be achieved by continuous motions that
maintain simplicity of the chain throughout, i.e.,
links are not permitted to intersect.
A chain that cannot be straightened or convexified we call {\em locked};
otherwise the chain is {\em unlocked}.
Note that a chain in 3D can be continuously moved between any of its
unlocked configurations, for example via straightened or convexified
intermediate configurations.

Basic questions concerning open and closed chains have proved surprisingly 
difficult.  
For example, the question of whether every planar, simple open chain 
can be straightened in the plane while maintaining simplicity 
has circulated in the computational geometry community for years, 
but remains open at this writing.
Whether locked chains exist in dimensions $d \ge 4$ was only settled
(negatively, in~\cite{co-pccl4d-99}) 
as a result of the open problem we posed in a preliminary
version of this paper~\cite{bddlloorstw-lupc3d-99}.
In piecewise linear knot theory, complete classification of the 3D 
embeddings of closed chains with $n$ edges has been found to be difficult, 
even for $n = 6$.  These types of questions are basic to the study of 
embedding and reconfiguration of edge-weighted graphs, where the weight 
assigned to an edge specifies the desired distance between the vertices 
it joins. Graph embedding and reconfiguration problems, with or without 
a simplicity requirement, have arisen in many contexts, 
including molecular conformation, 
mechanical design, 
robotics, 
animation,
rigidity theory, 
algebraic geometry,
random walks,
and knot theory.

We obtain several results for chains in 3D:
open chains with a simple orthogonal projection,
or embedded in the surface of a polytope,
may be straightened
(Sections~\secref{Open.3D} and~\secref{Open.Polytope});
but there exist open and closed chains that are locked
(Section~\secref{Knitting.Needles}).
For closed chains initially taking the form of a polygon lying in a plane,
it has long been known that
they may be convexified in 3D, but only via a procedure
that may require an unbounded number of moves.
We 
provide an algorithm to perform the
convexification
(Section~\secref{StLouis}) in $O(n)$ moves.

Previous computational geometry research on the reconfiguration
of chains 
(e.g., \cite{k-rpuras-97}, \cite{ksw-frit-96}, \cite{w-aigplm-92})
typically concerns planar chains with
crossing links, moving in the presence of obstacles; 
\cite{s-scsc-73}
and \cite{lw-rcpce-95}
reconfigure closed chains with crossing links in all dimensions $d \ge 2$.
In contrast, throughout this paper we work in 3D 
and
require that chains remain simple throughout their motions.  
Our algorithmic methods
complement the algebraic and topological approaches to these 
problems, offering constructive proofs for topological results and 
raising computational, complexity, and algorithmic issues.
Several open problems are listed in Section~\secref{Open}.

\conf{
The Schwartz-Sharir cell decomposition approach~\cite{ss-pmp2g-83}
from algorithmic robotics
shows that all the problems we consider in this paper are decidable,
and Canny's roadmap algorithm~\cite{c-crmp-87} leads to
an algorithm singly-exponential
in $n$.
See, e.g.,
\cite{hjw-mp2dl-84},
\cite{k-gir-85}, 
\cite{ch-dgmc-88}, 
or
\cite{w-rsa-97}
for other weighted graph embedding and reconfiguration problems.
}

\subsection{Background}
Thinking about movements of polygonal chains goes back at least
to A.~Cauchy's 1813 theorem on the rigidity of polyhedra~\cite[Ch.~6]{c-p-97}.
His proof employed a key lemma on opening angles at the joints of
a planar convex open polygonal chain.
This lemma, now known as Steinitz's Lemma (because E.~Steinitz gave
the first correct proof in the 1930's), is similar in spirit to
our Lemma~\lemref{barb}.
Planar linkages, objects more general than polygonal chains in
that a graph structure is permitted, have been studied
intensively by mechanical engineers since at least Peaucellier's 1864 linkage.
Because the goals of this linkage work are so different from ours,
we could not find directly relevant results in the literature
(e.g., \cite{h-kgm-78}).  However, we have no doubt that simple
results like our convexification of quadrilaterals (Lemma~\lemref{quad.M02})
are known to that community.

Work in algorithmic robotics is relevant.
In particular, the Schwartz-Sharir cell decomposition approach~\cite{ss-pmp2g-83}
shows that all the problems we consider in this paper are decidable,
and Canny's roadmap algorithm~\cite{c-crmp-87} leads to
an algorithm singly-exponential in $n$, the number of vertices of the
polygonal chain.
Although hardness results are known for more general 
linkages~\cite{hjw-mp2dl-84},
we know of no nontrivial lower bounds for the problems discussed in
this paper.

See, e.g.,
\cite{hjw-mp2dl-84},
\cite{k-gir-85}, 
\cite{ch-dgmc-88}, 
or
\cite{w-rsa-97}
for other weighted graph embedding and reconfiguration problems.

\subsection{Measuring Complexity}
\seclab{Complexity}
As usual, we compute the time and space 
complexity of our algorithms as a function of
$n$, the number of vertices of the polygonal chain.
This, however, will not be our focus, for
it is of perhaps greater interest 
to measure the geometric complexity
of a proposed reconfiguration of a chain.
We first define what constitutes a ``move'' for these
counting purposes.

Define a {\em joint movement\/} at $v_i$ to be a monotonic rotation
of $e_i$ about an axis through $v_i$ fixed with respect to a reference frame
rigidly attached to some other edges of the chain.
For example, 
a joint movement could feasibly be executed by a motor at $v_i$
mounted in a frame attached to $e_{i-1}$ and $e_{i-2}$.
The axis might be moving in absolute space (due to other joint movements), 
but it must be fixed in the reference frame.
Although more general movements could be explored, these will
suffice for our purposes.
A {\em monotonic rotation\/} does not stop or reverse direction.
Note we ignore the angular velocity profile of a joint movement,
which might not be appropriate in some applications.
Our primary measure of complexity is a {\em move\/}:
a reconfiguration of the chain $P$ of $n$ links to
$P'$ that may be composed of a constant number of simultaneous
joint movements.  Here the constant number should be independent
of $n$, and is small ($\le 4$)
in our algorithms.
All of our algorithms achieve reconfiguration in $O(n)$ moves.
One of our open problems (Section~\secref{Open}) asks for
exploration of a measure of the complexity of movements.

\section{Open Chains with Simple Projections}
\seclab{Open.3D}
This section considers an open polygonal chain $P$ in 3D with a simple
orthogonal projection $P'$ onto a plane.
Note that there is a polynomial-time
algorithm to determine whether $P$ admits such a projection, 
and to output a projection plane if it exists~\cite{bgrt-dnpos-96}.
We choose our coordinate system so that 
the $xy$-plane $\pixy$ is parallel to this plane;
we will refer to lines and planes parallel to the $z$-axis as ``vertical.''
We will describe an algorithm that straightens $P$, working from
one end of the chain.
We use the notation
$P[i,j]$ to represent the chain of edges 
$(v_i, v_{i+1},\ldots,v_j)$, including $v_i$ and $v_j$,
and $P(i,j)$ to represent the chain without its endpoints: 
$P(i,j) = P[i,j] \setminus \{ v_i , v_j \}$.
Any object lying in plane
$\pixy$ will be labelled with a prime.

Consider the projection $P'=(v'_0,v'_1,\ldots,v'_{n-1})$ on $\pixy$.
Let $r_i = \min_{j\not\in \{i-1,i\}} d(v'_i,e'_j)$, where
$d(v',e')$ is the minimum distance from vertex $v'$ to a point on edge
$e'$.  Construct a disk of radius $r_i$ around each vertex $v'_i$.
The interior of each disk
does not intersect any other vertex
of $P'$ and does not intersect any edges other than the two incident
to $v'_i$: $e'_{i-1}$ and $e'_i$; see Fig.~\figref{simple.proj}.
\conf{
to $v'_i$: $e'_{i-1}$ and $e'_i$.
}
\begin{figure}[htbp]
\begin{center}
\ \psfig{figure=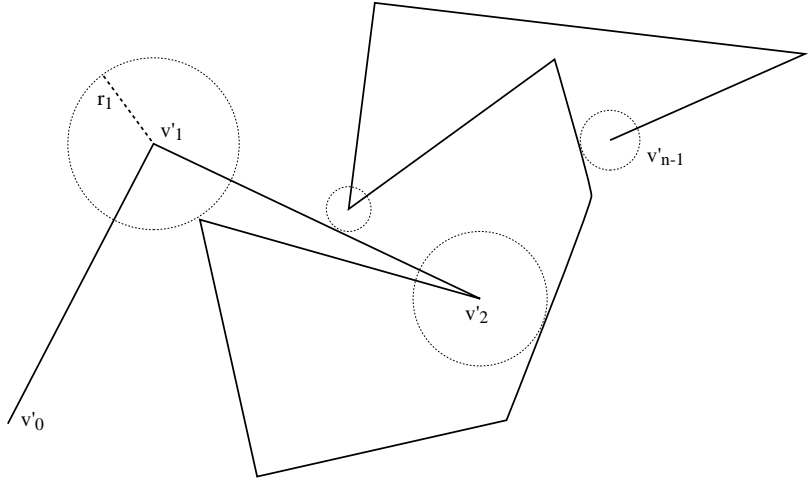,height=2.5in}
\conf{
\ \psfig{figure=simple.proj.eps,width=8cm}
}
\end{center}
\caption{The projection $P'$ of $P$.  Each vertex $v'_i$ is surrounded by
an ``empty'' disk of radius $r_i$. Several such disks are shown.}
\figlab{simple.proj}
\end{figure}

We construct in 3D a vertical cylinder $C_i$ centered on each vertex
$v_i$ of radius $r = \frac{1}{3}\min_i \{ r_i \}$.  This choice of $r$
ensures that no two cylinders intersect one another (the choice of the
fraction $\frac{1}{3} < \frac{1}{2}$ guarantees that cylinders do not even touch),
and no edges of $P$, other than those incident to $v_i$, intersect
$C_i$, for all $i$.  

The straightening algorithm proceeds in two stages.
In the first stage, the links are squeezed like an accordion
into the cylinders, so that after step $i$ all the links
of $P_{i+1}=P[0,i+1]$ are packed into $C_{i+1}$.
Let $\Pi_i$ be the vertical plane containing
$e_i$ (and therefore $e'_i$).
After the first stage, the chain is {\em monotone\/} in $\Pi_i$,
i.e., it is monotone with respect to the line $\Pi_i \cap \Pi_{xy}$
in that the intersection of the chain with a vertical line in $\Pi_i$ is either
empty or a single point.
In stage two, the chain
is unraveled link by link
into a straight line.
The rest of this section describes the first stage.
Let $\d = r/n$.

\subsection{Stage 1}
We describe the Step~0 and the general Step~$i$ separately,
although the former is a special case of the latter. 
\begin{enumerate}
\conf{\squeezelist}
\item[0.]
\conf{\squeezelist}
\begin{enumerate}
\conf{\squeezelist}
\item Rotate $e_0$ about $v_1$, within $\Pi_0$, 
so that the projection of $e_0$
on $\pixy$ is contained in $e'_0$ throughout the motion.
The direction of rotation is determined by the relative heights
($z$-coordinates) of $v_0$ and $v_1$.
Thus
if $v_0$ is at or above $v_1$, $e_0$ is rotated upwards
($v_0$ remains above $v_1$ during the rotation);
see Fig.~\figref{cylinder}.
If $v_0$ is lower than $v_1$, $e_0$ is rotated downwards
($v_0$ remains below $v_1$ during the rotation).
The rotation stops when $v_0$ lies within $\d$ of the vertical
line through $v_1$, i.e., when $v_0$ lies in the cylinder $C_1$
and is very close to its axis.
The value of $\d$ is chosen to be $r/n$ so that in later steps
more links can be accommodated in the cylinder.
Again see Fig.~\figref{cylinder}.
\item
Now we rotate $e_0$ about the axis of $C_1$ away from $e_1$,
until $e'_0$ and $e'_1$
are collinear (but not overlapping), 
i.e., until $e_0$ lies in the vertical plane $\Pi_1$. 
\end{enumerate}
\conf{
See~Fig.~\figref{cylinder}.
}
\begin{figure}[htbp]
\begin{center}
\ \psfig{figure=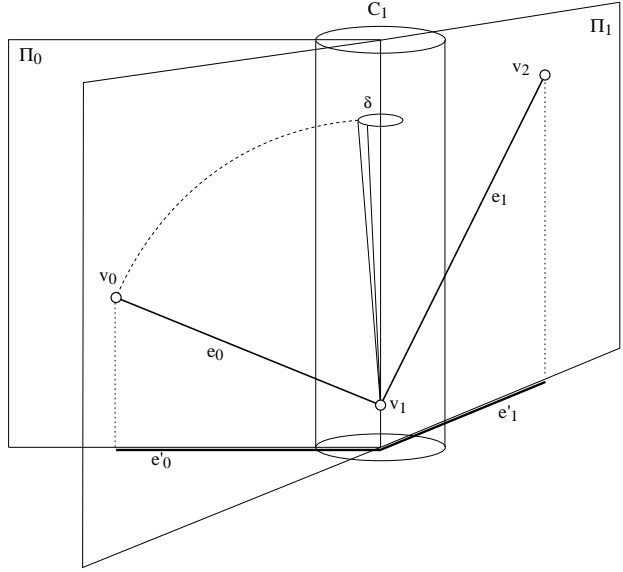,height=3in}
\conf{
\ \psfig{figure=cylinder.eps,width=7cm}
}
\end{center}
\caption{
	Step~0:  (a) $e_0$ is first rotated within $\Pi_0$ into $C_1$,
	and then (b) rotated into the vertical plane $\Pi_1$ containing $e_1$.
\conf{
	Step~0:  $e_0$ is rotated within $\Pi_0$ and then into $\Pi_1$.
}
}
\figlab{cylinder}
\end{figure}
After completion of Step~0, $(v_0, v_1, v_2)$ forms a 
chain in $\Pi_1$ monotone with respect to the line $\Pi_1 \cap \pixy$.
\item[$i$.]
At the start of Step~$i>0$, we 
have a monotone chain 
$P_{i+1} = P[0,i+1]$
contained in the vertical plane $\Pi_i$ through $e_i$, 
with $P_i=P[0,i]$ in $C_i$ and $v_0$ within a distance of $i\d$
of the axis of $C_i$.
\begin{enumerate}
\conf{\squeezelist}
\item
As in Step~0(a), rotate
\conf{
Rotate
}
$e_i$ within $\Pi_i$
(in the direction that shortens the vertical projection of $e_i$)
so that
$v_i$ lies within a distance $\d$ of the axis of $C_{i+1}$.
The difference now is that $v_i$ is not the start of the chain,
but rather is connected to the chain $P_i$.
During the rotation of $e_i$ we ``drag'' $P_i$ along in such a way that
only joints $v_i$ and $v_{i+1}$ rotate, keeping
the joints $v_1,\ldots,v_{i-1}$ frozen.
Furthermore, we constrain the motion of $P_i$ 
(by appropriate rotation about joint $v_i$)
so that it does
not undergo a rotation. Thus at any instant of time during the
rotation of $e_i$, the position of $P_i$ remains within $\Pi_i$
and is a translated copy of the initial $P_i$.
See Fig.~\figref{accordion}.
\begin{figure}[htbp]
\begin{center}
\ \psfig{figure=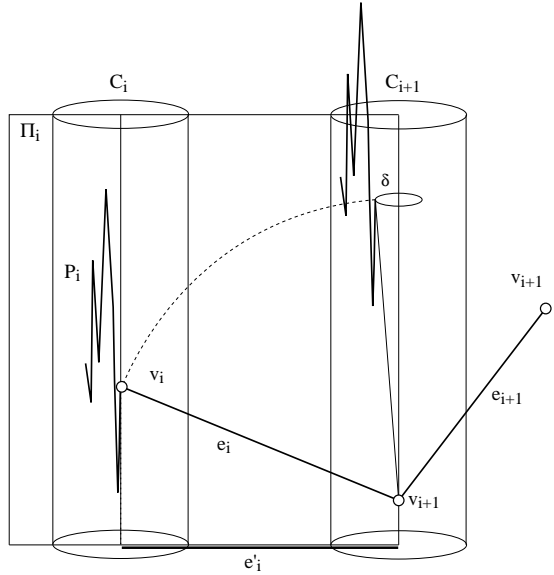,height=3in}
\conf{
\ \psfig{figure=accordion.eps,height=6.5cm}
}
\end{center}
\caption{The chain $P_i$ translates within $\Pi_i$.}
\figlab{accordion}
\end{figure}
\item
Following Step~0(b), rotate 
\conf{
Rotate 
}
$P_{i+1}$ 
about the axis of $C_{i+1}$ until $e'_i$ and $e'_{i+1}$ are coplanar.
\end{enumerate}
At the completion of Step~$i$ we therefore have a chain
$P_{i+2} =P[0,i+2]$ in the vertical plane $\Pi_{i+1}$, with
$P_{i+1}$ in $C_{i+1}$ and $v_0$ within a distance of $(i+1)\d$ of its axis.
The chain is monotone in $\Pi_{i+1}$
with respect to the line $\Pi_{i+1} \cap \pixy$.
\end{enumerate}
\conf{
Now the second stage can be performed simply by straightening
one joint at a time, because this operation maintains monotonicity.
}

\subsection{Stage 2}
Now it is trivial to unfold this monotone chain by straightening one
joint at a time, i.e., rotating each joint angle to $\pi$, 
starting at either end of the chain.
We have therefore established the first claim of this theorem:

\begin{theorem}
A polygonal chain of $n$ links
with a simple orthogonal projection may be straightened,
in $O(n)$ moves, with an algorithm of
$O(n)$ time and space complexity.
\theolab{simple.proj}
\end{theorem}

Counting the number of moves is straightforward.
Stage~1, Step~$i$(a) requires one move:
only joints $v_i$ and $\v1$ rotate.
Step~$i$(b) is again one move: only $\v1$ rotates.
So Stage~1 is completed in $2n$ moves.
As Stage~2 takes $n-1$ moves, the whole procedure is accomplished
with $O(n)$ moves.

Each move can be computed in constant time, so the time complexity
is dominated by the computation of the cylinder radii $r_i$.
These can be trivially computed in $O(n^2)$ time, by computing
each vertex-vertex and vertex-edge distance.
However, a more efficient computation is possible, based on the
medial axis of a polygon, as follows.
Given the projected chain $P'$ in the plane 
(Fig.~\figref{medial}a), form two simple polygons
$P_1$ and $P_2$, by doubling the chain from its endpoint
$v'_0$ until the convex hull is reached (say at point $x$),
and from there connecting along the line bisecting the
hull angle at $x$ to a large surrounding rectangle, and similarly
connecting from $v'_{n-1}$ to the hull to the rectangle.
For $P_1$ close the polygon above $P'$, and below for $P_2$.
See Figs.~\figref{medial}bc. Note that $P_1 \cup P_2$ covers
the rectangle, which, if chosen large, effectively covers the plane
for the purposes of distance computation.
\begin{figure}[htbp]
\begin{center}
\ \psfig{figure=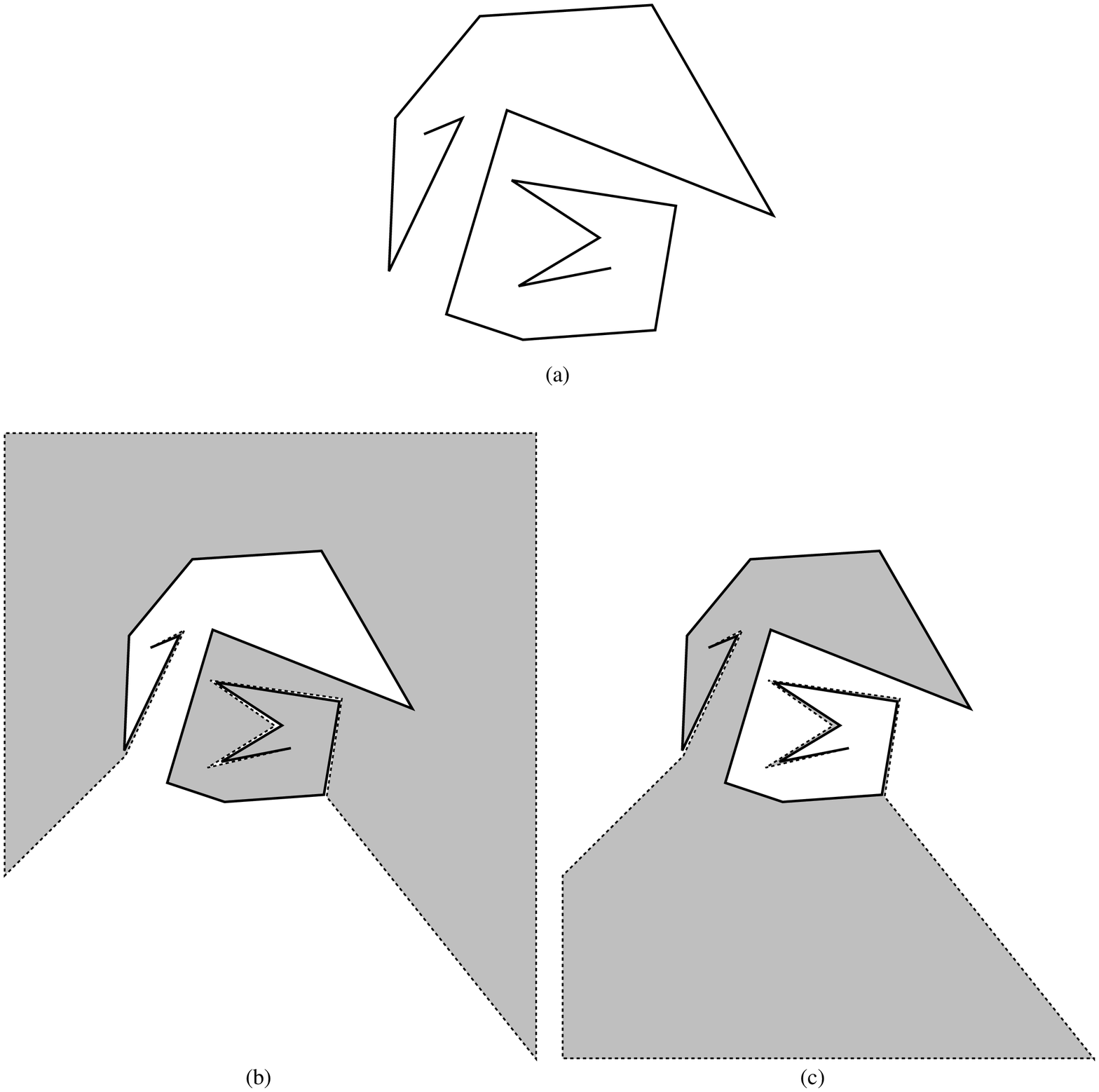,height=12cm}
\end{center}
\caption{(a) Chain $P'$; (b) Polygon $P_1$; (c) Polygon $P_2$.}
\figlab{medial}
\end{figure}

Compute the medial axis of $P_1$ and $P_2$ using a
the linear-time algorithm of~\cite{csw-fmasp-95}.
The distances $r_i$ can now be determined from the
distance information in the medial axes.
For a convex vertex $v_i$ of $P_k$, its minimum
``feature distance'' can be found from axis information
at the junction of the axis edge incident to $v_i$.
For a reflex vertex, the information is with the associated
axis parabolic arc.
Because the bounding box is chosen to be large, no
vertex's closest feature is part of the bounding box,
and so must be part of the chain.

\section{Open Chains on a Polytope}
\seclab{Open.Polytope}
In this section we show that any open chain embedded on the
surface of a convex polytope may be straightened.  We start with a
planar chain which we straighten in 3D.

Let $P$ be an open chain in 2D, lying in $\pixy$.
It may be easily straightened by the following procedure.
Rotate $e_0$ within $\Pi_0$ until it is vertical; now
$v_0$ projects into $v_1$ on $\pixy$.
In general, rotate $e_i$ within $\Pi_i$ until $v_i$
sits vertically above $v_{i+1}$.
Throughout this motion, keep the previously straightened 
chain $P_i=P[0,i]$ above $v_i$ in a vertical ray through $v_i$.
This process clearly maintains simplicity throughout,
as the projection at any stage is a subset of the original
simple chain in $\pixy$.
In fact, this procedure can be seen as a special case of
the algorithm described in the preceding section.

An easy generalization of this ``pick-up into a vertical ray''
idea permits straightening any open chain lying on the
surface of a convex polytope $\P$.
The same procedure is followed, except that the 
surface of $\P$ plays the role of $\pixy$, and surface
normals play the roles of vertical rays.
When a vertex $v_i$ of the polygonal chain $P$ lies on
an edge $e$ between two faces $f_1$ and $f_2$ of $\P$,
then the line containing $P_i$ is rotated from $R_1$,
the ray through $v_i$ and normal to $f_1$, through 
an angle of measure $\pi - \d(e)$,
where $\d(e)$ is the (interior) dihedral angle at $e$,
to $R_2$, 
the ray through $v_i$ and normal to $f_2$.

This algorithm uses $O(n)$ moves and can be executed in $O(n)$ time.

Note that it is possible to draw a polygonal chain on a polytope
surface that has no simple projection.
So this algorithm handles some cases not covered by 
Theorem~\theoref{simple.proj}.
We believe that the sketched algorithm applies to
a class of polyhedra
wider than convex polytopes, 
but we will not pursue this further here.
\conf{
Any open chain lying on the surface of a convex polytope
can be straightened
by ``picking up'' the chain.  Working from one end of the chain, each link is
rotated up away from the polytope to the normal of the face
containing the link.  At edges of the polytope the straightened subchain
is swiveled from one
face normal to the next.  For details see~\cite{full}.
Though related to the result of the previous section, this result is
different because a polygonal 
chain on a polytope surface need not have a simple orthogonal projection.
}

\section{Locked Chains}
\seclab{Knitting.Needles}
Having established that two classes of open chains may be
straightened, we show in this section that not all open chains
may be straightened, describing one locked open chain of
five links (Section~\secref{Locked.Open}).  
A modification of this example establishes
the same result for closed chains (Section~\secref{Locked.Closed}).
Both of these results were obtained independently by other 
researchers~\cite{cj-nepiu-98}.
Our proofs are, however, sufficiently different
to be of independent interest.
\subsection{A Locked Open Chain}
\seclab{Locked.Open}
Consider the chain $K=(v_0,\ldots,v_5)$ configured as in
Fig.~\figref{knitting},
where the standard knot theory convention is followed to denote
``over'' and ``under'' relations.
Let $L = \ell_1 + \ell_2 + \ell_3$ be the total length of the
short central links, and
let $\ell_0$ and $\ell_4$ be both larger than $L$;
in particular, choose $\ell_0 = L + \d$ and $\ell_4 = 2L + \d$ for $\d > 0$.
(One can think of this as 
composed of two rigid knitting needles, $e_0$
and $e_4$, connected by a flexible cord of length $L$.)
Finally, center a ball $B$ of radius $r = L + \e$ on $v_1$,
with $0 < 2\e < \d$.
The two vertices $v_0$ and $v_5$ are exterior to $B$, while the
other four are inside $B$.
See Fig.~\figref{knitting}.
\begin{figure}[htbp]
\begin{center}
\ \psfig{figure=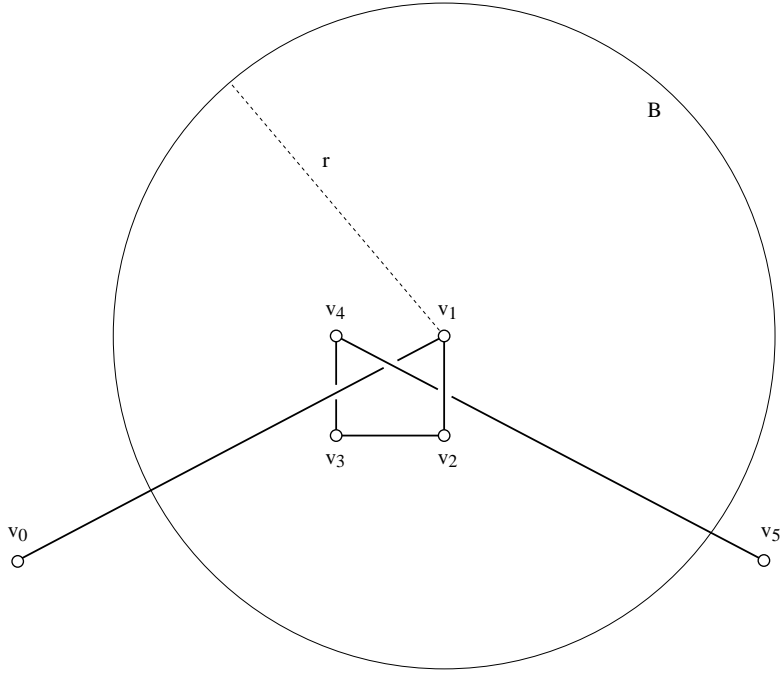,height=3.5in}
\conf{
\ \psfig{figure=knitting.eps,width=7.5cm}
}
\end{center}
\caption{A locked open chain $K$ (``knitting needles'').
(The first and last edges $e_0$ and $e_4$ are longer than
they appear in this view.)}
\figlab{knitting}
\end{figure}

Assume now that the chain $K$ can be straightened by some motion.
During the entire process,
$\{ v_1, v_2, v_3, v_4 \} \subset B$ because $L < r$.
Of course $v_0$ remains outside of $B$ because $\ell_0 > r$.
Now because $v_4 \in B$ and
$\ell_4 = |v_4v_5| = 2L+\d$ is more than the diameter $2r = 2(L+\e)$ of $B$,
$v_5$ also remains exterior to $B$ throughout the motion.

Before proceeding with the proof, we recall some terms from
knot theory.
The {\em trivial knot\/} is an unknotted closed curve homeomorphic
to a circle.  
The {\em trefoil knot\/} is the simplest
knot, the only knot that may be drawn with three crossings.
See, e.g.,~\cite{Livingston} or~\cite{a-kb-94}.
\conf{
See, e.g.,~\cite{a-kb-94}.
}

Because of the constant separation between
$\{ v_0, v_5 \}$ and
$\{ v_1, v_2, v_3, v_4 \}$
by the boundary of $B$, 
we could have attached
a sufficiently long unknotted string $P'$ from $v_0$ to $v_5$
exterior to $B$ that would not have hindered the unfolding
of $P$.  
But this would imply that $K \cup P'$ is the
trivial knot; but it is clearly a trefoil knot.
We have reached a contradiction; therefore, $K$ cannot be straightened.

\subsection{A Locked, Unknotted Closed Chain}
\seclab{Locked.Closed}
It is easy to obtain locked closed chains in 3D: simply tie the
polygonal chain into a knot.  Convexifying such a chain would
transform it to the trivial knot, an impossibility.
More interesting for our goals is whether there exists a
locked, closed polygonal chain that is {\em unknotted}, 
i.e., whose topologically
structure is that of the trivial knot.

We achieve this by ``doubling'' $K$: adding
vertices $v'_i$ near $v_i$ for $i=1,2,3,4$, and connecting
the whole into a chain $K^2 = (v_0,\ldots,v_5, v'_4,\ldots,v'_1)$.
See Fig.~\figref{knitting2}.
\begin{figure}[htbp]
\begin{center}
\ \psfig{figure=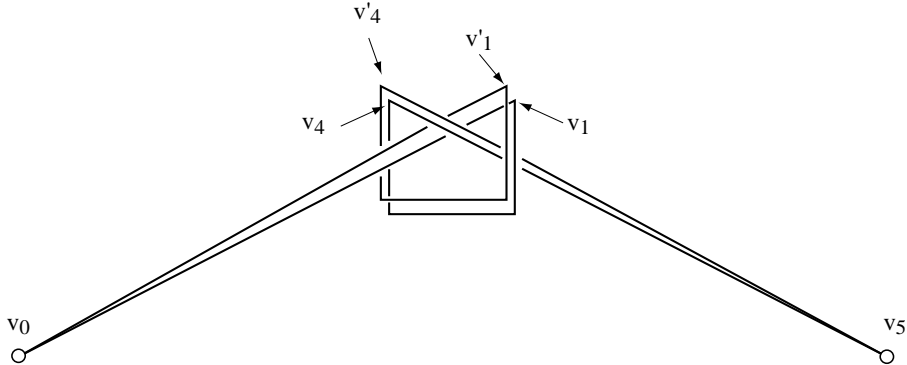,width=12cm}
\conf{
\ \psfig{figure=knitting2.eps,width=8.5cm}
}
\end{center}
\caption{$K^2$ ($K$ doubled): a locked but unknotted chain.}
\figlab{knitting2}
\end{figure}
Because $K \subset K^2$, the preceding argument applies when
the second copy of $K$ is ignored:
any convexifying motion will have the property that $v_0$ and $v_5$
remain exterior to $B$, and
$\{ v_1, v_2, v_3, v_4 \}$ remain interior to $B$ throughout
the motion.  
Thus the extra copy of $K$ provides
no additional freedom of motion to $v_5$ with respect to $B$.
Consequently, we can argue as before:  if $K^2$ is somehow
convexified, this motion could be used to unknot $K \cup P'$,
where $P'$ is an unknotted chain exterior to $B$ connecting $v_0$ to $v_5$.
This is impossible, therefore $K^2$ is locked.

\section{Convexifying a Planar Simple Polygon in 3D}
\seclab{StLouis}
An interesting open problem is to generalize our result from Section~\secref{Open.3D}
to convexify a general closed chain.
We show now that the special case of a closed chain
lying in a plane, i.e., a planar simple polygon, 
may be convexified in 3D.

Such a polygon may be
convexified in 3D by ``flipping'' out the reflex pockets,
i.e., rotating the pocket chain into 3D and back down to the plane;
see Fig.~\figref{pocket}.
This simple procedure was suggested by Erd\H{o}s~\cite{e-p3763-35}
and proved to work by de Sz.~Nagy~\cite{sn-sp3763-39}.
The number of flips, however, cannot be bound as a function
of the number of vertices $n$ of the polygon, as
first proved by Joss and Shannon~\cite{g-hcp-95}.
See~\cite{t-entr-99} for the complex history of these results.
\begin{figure}[htbp]
\begin{center}
\ \psfig{figure=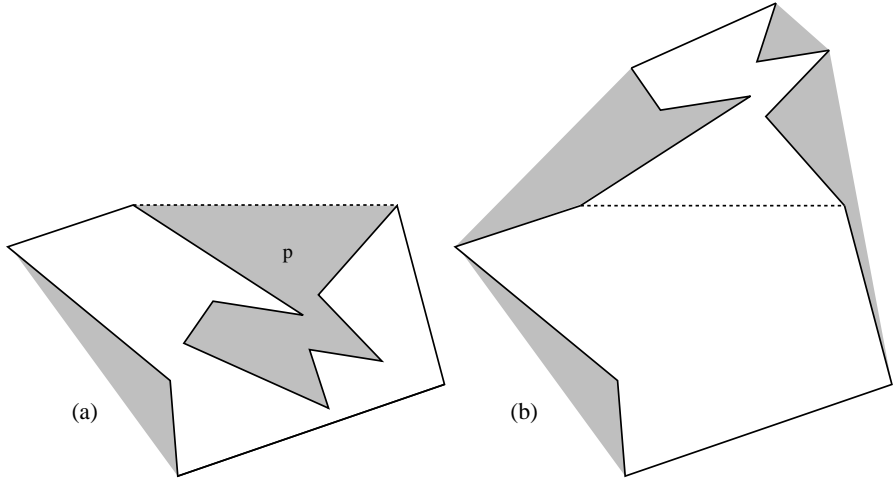,height=2.5in}
\end{center}
\caption{(a) A pocket $p$; (b) The polygon after flipping $p$.}
\figlab{pocket}
\end{figure}

We offer a new algorithm for convexifying planar closed chains,
which we call the ``St. Louis Arch'' algorithm.
It is more complicated than flipping
but uses a bounded number of moves, in fact $O(n)$ moves.
It models the intuitive approach of picking up the polygon
into 3D.  We discretize this to lifting vertices one by one, accumulating
the attached links into a convex
``arch''\footnote{
        We call this the {\em St.\ Louis Arch Algorithm\/}
        because of the resemblance to the arch in St.\ Louis, Missouri.
}
$A$
in a
vertical plane above the remaining polygonal chain;
see Fig.~\figref{A0}.
Although the
algorithm is conceptually simple, some care is required to make it
precise, and to then establish that simplicity is maintained
throughout the motions.

Let $P$ be a simple polygon in the $xy$-plane, $\pixy$.
Let $\pie$ be the plane $z = \e$ parallel to $\pixy$, for $\e > 0$;
the value of $\e$ will be specified later.
We use this plane to convexify the arch safely above the
portion of the polygon not yet picked up.
We will use primes to indicate positions of moved (raised) vertices;
unprimed labels refer to the original positions.
After a generic step $i$ of the algorithm,
$P(0,i)$ has been lifted above $\pie$ and convexified,
$v_0$ and $v_i$ have been raised to $v'_0$ and $v'_i$ on $\pie$,
and
$P[i+1,n-1]$ remains in its original
position on $\pixy$.
We first give a precise description of the conditions that
hold after the $i$th step.
Let $\Pi_z(v_i,v_j)$ be the (vertical) plane containing 
\conf{ 
$v_i$ and $v_j$.
} 
$v_i$ and $v_j$,
parallel to the $z$-axis.

\begin{enumerate}
\conf{\squeezelist}
\item[H1:]
$\pie$ splits the vertices of $P$ into three sets: $v'_0$ and $v'_i$
lie in $\pie$, $v'_1, \ldots, v'_{i-1}$ lie above the plane, and
$v_{i+1}, \ldots, v_{n-1}$ lie below it.
\item[H2:]
The arch $A = P(0,i)$ lies in the plane $\Pi_z(v'_0, v'_i)$, and is convex.
\item[H3:] $v'_0$ and $v'_i$ project onto $\pixy$ within distance $\d$ of
their original positions $v_0$ and $v_i$.
(Here, $\delta>0$ is
a constant that depends only on the input positions; it will
be specified later.)

\item[H4:]
Edges $\vn1 v'_0$ and $v'_i \v1$ connect between
$\pixy$ and $\pie$.
\item[H5:]
$P[i+1,n-1]$ remains in its original
position in $\pixy$.
\end{enumerate}
See Fig.~\figref{A0}.
\begin{figure}[htbp]
\begin{center}
\ \psfig{figure=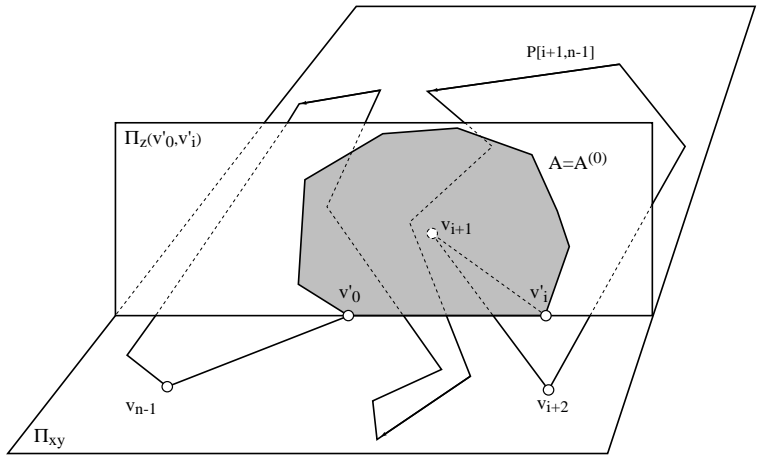,width=10cm}
\conf{
\ \psfig{figure=A0.eps,width=8cm}
}
\end{center}
\caption{The arch $A$ after the $i$th step, i.e.,
after ``picking up'' $P(0,i)$ into $A$.
(The planes $\pixy$ and $\pie$ are not distinguished in
this figure, nor in Figs.~\protect\figref{A1} or~\protect\figref{A2}.)
}
\figlab{A0}
\end{figure}
A central aspect of the algorithm will be choosing $\e$ small
enough to guarantee a $\d$ (see H3) that maintains simplicity
throughout all movements.

The algorithm consists of an initialization step S0, followed by 
repetition of steps S1--S4.

\subsection{S0}
The algorithm is initialized at $i=2$ by selecting an arbitrary
(strictly) convex vertex $v_1$, 
and raising $\{ v_0, v_1, v_2 \}$ in four 
steps:
\begin{enumerate}
\conf{\squeezelist}
\item Rotate $v_1$ about the line through $v_0 v_2$ up to $\pie$.
Call its new position $v''_1$.
\item Rotate $v_0$ about the line through $v_{n-1} v''_1$ up to $\pie$.
Call its new position $v'_0$.
\item Rotate $v_2$ about the line through $v''_1 v_3$ up to $\pie$.
Call its new position $v'_2$.
\item Rotate $v''_1$ about the line through $v'_0 v'_2$ upwards
until it lies in the plane $\Pi_z(v'_0,v'_2)$.
Call its new position $v'_1$.
\end{enumerate}
So long as the joint at $v''_1$ is not straight,
the $4$th step above is unproblematical, simply rotating a triangle
from a horizontal to a vertical plane.
That this joint does not become straight
depends on $\e$ and $\d$, and will be established
under the discussion of S1 below.  Ditto for establishing that
the first three steps can be accomplished without causing
self-intersection.
\conf{
steps, whose details we leave for~\cite{full}.
}

After completion of Step S0, the hypotheses H1--H5 are all satisfied.
The remaining steps S1--S4 are repeated for each $i > 2$.

\subsection{S1}
The purpose of Step S1 is to lift $v_{i}$ from $\pixy$ to $\pie$.
This will be accomplished by a rotation of $v_{i}$ about
the line through $v'_{i-1}$ and 
\conf{
$v_{i+1}$.
}
$v_{i+1}$, the same rotation used
in substeps~(2) and~(3), and in a modified form in~(1), of Step S0.
Although this rotation is conceptually simple, it is this key movement
that demands a value of $\e$ to guarantee a $\d$ that ensures correctness.
The values of $\e$  and $\d$ will be computed directly from the 
initial geometric structure of $P$.  
Specifying the conditions on $\e$ is one of the more delicate
aspects of our argument, to which we now turn.

Let $\a_j$ be the smaller of the two (interior and exterior)
angles at $v_j$.  Also let $\b_j = \pi - \a_j$, the deviation
from straightness at joint $v_j$.
We assume that $P$ has no three consecutive collinear vertices.
If a vertex is collinear with its two adjacent vertices,
we freeze and eliminate that joint.
So we may assume that $\b_j > 0$ for all $j$.

\subsubsection{Determination of $\d$}
\seclab{delta}
As in our earlier Figure~\figref{simple.proj}, the simplicity of $P$
guarantees ``empty'' disks around each vertex.  Here we need
disks to meet more stringent conditions than used
in Section~\secref{Open.3D}.
Let $\d > 0$ be such that:
\begin{enumerate}
\conf{\squeezelist}
\item Disks around each vertex $v_j$ of radius $\d$ include no
other vertices of $P$, and only intersect the two edges incident
to $v_j$.
\item 
A perturbed polygon, obtained by displacing the vertices within
the disks (ignoring the fixed link lengths),
   \begin{enumerate}
   \item remains simple, and
   \item has no straight vertices.
   \end{enumerate}
\end{enumerate}

It should be clear that the simplicity of $P$ together with $\b_j > 0$
guarantees that such a $\d > 0$ exists.  
As a technical aside,
we sketch how $\d$ could be computed.
Finding a radius that satisfies condition~(1) is easy.
Half this radius guarantees the simplicity condition~(2a),
for this keeps a maximally displaced vertex separated from a
maximally displaced edge.
To prevent an angle $\b_j$ from reaching zero,
condition~(2b),
displacements of
the three points $v_{j-1}$, $v_j$, and $v_{j+1}$ must be
considered.  
Let $\ell = \min_j \{ \ell_j \}$ be the length of the shortest
edge, and let $\b' = \min_j \{ \b_j \}$ 
be the minimum deviation from collinearity.
Lemma~\lemref{rho},which we prove in the Appendix, shows that
choosing
$\d < \frac{1}{2} \ell \sin (\b'/2)$
prevents straight vertices.

Let $\s$ be the minimum separation
$|v_j v_k|$ for all positions of $v_j$ and $v_k$ within
their $\d$ disks, for all $j$ and $k$.  
Condition~(2a) guarantees that $\s > 0$.
Note that $\s \le \ell$.
Let $\b$ be the minimum of all $\b_j$ for all positions of $v_j$ within
their $\d$ disks.  
Condition~(2b) guarantees that $\b > 0$.
Our next task is to derive $\e$ from $\s$, $\b$, and $\d$.
To this end, we must detail the ``lifting'' step of the algorithm.
\conf{
First, we sketch the computation of $\d$.  It is chosen so that
disks of radius $\d$ around each vertex are empty of other vertices
and all but the two incident edges, and, most importantly, that
displacement of the vertices within these disks (perhaps all
simultaneously) both maintains polygon simplicity and could not
align any three vertices to become collinear.
}

\subsubsection{S1 Lifting}
\seclab{S1.lifting}
Throughout the algorithm, $v'_0$ remains fixed at the position on $\pie$
it reached  in Step S0.
During the lifting step, $v'_{i-1}$ also remains fixed, while $v_i$ is lifted.
Thus $v'_0 v'_{i-1}$, the base of the arch $A$, remains fixed during the
lifting, which permits us, by hypothesis H1, to safely ignore
the arch during this step.

We now concentrate on the $2$-link chain $(v'_{i-1}, v_i, v_{i+1})$.
By H5, $v_i v_{i+1}$ has not moved on $\pixy$;
by H3, $v'_{i-1}$ has not moved horizontally more than $\d$ from $v_{i-1}$.
Let $\a'_{i}$ be the measure in $[0,\pi]$ of
angle $\angle(v'_{i-1}, v_i, v_{i+1})$, i.e., the angle at $v_i$ measured
in the slanted plane determined by the three points.
Because $v_i v_{i+1}$ lie on $\pixy$ and $v'_{i-1}$ is on $\pie$,
$\a'_{i} \neq \pi$ and
the chain $(v'_{i-1}, v_i, v_{i+1})$ is kinked at the joint $v_i$.

Now imagine holding $v'_{i-1}$ and $v_{i+1}$ fixed.
Then $v_i$ is free to move on a circle $C$ with center on $v'_{i-1} v_{i+1}$.
See Fig.~\figref{double.cone}.
This circle might lie partially below $\pixy$,
and is tilted from the vertical (because $v'_{i-1}$ lies on $\pie$).
\begin{figure}[htbp]
\begin{center}
\ \psfig{figure=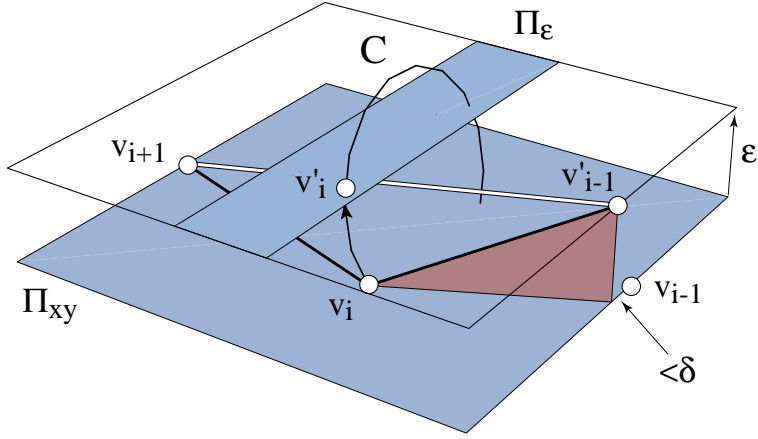,width=10cm}
\end{center}
\caption{$v_{i}$ rotates up the circle $C$ until it hits $\pie$.}
\figlab{double.cone}
\end{figure}
The lifting step consists simply in rotating $v_i$ on $C$ upward
until it lies on $\pie$; its position there we call $v'_i$.

\subsubsection{Determination of $\e$}
\seclab{Epsilon}
We now choose $\e>0$ so that two conditions are satisfied:
\begin{enumerate}
\conf{\squeezelist}
\item The highest point of $C$ is above $\pie$
(so that $v_i$ can reach $\pie$).
\item $v'_i$ projects no more than $\d$ away from $v_i$
(to satisfy H3).
\end{enumerate}
It should be clear that both goals may be achieved by choosing
$\e$ small enough.
We sketch a computation of $\e$ in the Appendix.

The computation of $\e$ 
ultimately depends solely on $\s$ and $\b$---the
shortest vertex separation 
and the smallest deviation from straightness---because
these determine $\d$, and then $r$ and 
$\d_1$ and $\d_2$ and $\e$.
\conf{
The computation of $\e$
ultimately depends solely on the
shortest vertex separation
and the smallest deviation from straightness.
}
Although we have described the computation within
Step S1, in fact it is 
\conf{
Thus it can be
}
performed prior to starting any movements;
and $\e$ remains fixed throughout.

As we mentioned earlier, two of the three lifting rotations used
in Step~S0 match the lifting just detailed.
The exception is the first lifting, of $v_1$ to $v'_1$ in Step~S0.
This only differs in that the cone axis $v_{0} v_2$ lies
on $\pixy$ rather than connecting $\pixy$ to $\pie$.  
But it should be clear this only changes the above
computation in that the tilt angle $\psi$ is zero, 
which only improves the inequalities.  Thus the $\e$ computed
for the general situation already suffices for this special case.

\subsubsection{Collinearity}
\seclab{Collinearity}
We mention here, for reference in the following steps,
that it is possible that $v'_i$ might be collinear with
$v'_0$ and $v'_{i-1}$ on $\pie$.
There are two possible orderings of these three vertices
along a line:
\begin{enumerate}
\conf{\squeezelist}
\item $(v'_0, v'_i ,v'_{i-1})$.
\item $(v'_0, v'_{i-1}, v'_i)$.
\end{enumerate}
The ordering
$(v'_{i}, v'_0, v'_{i-1})$
is not possible 
because that would violate the simplicity condition 2(a),
as all three vertices project to within $\d$ of their
original positions on $\pixy$, and no vertex comes within
$\delta$ of an edge.

Despite this possible degeneracy, we will refer to ``the
triangle $\triangle v'_0 v'_{i-1} v'_i$,'' with the understanding that
it may be degenerate.
This possibility will be dealt with in Lemma~\lemref{strict}.

We now turn to the remaining three steps of the algorithm for
iteration $i$.
We use the 
notation $A^{(k)}$ to represent
the arch $A=A^{(0)}$ at various stages of its processing, incrementing
$k$ whenever the shape of the arch might change.

\subsection{S2}
After the completion of Step~S1, $v'_{i-1} v'_i$ lies in $\pie$.
We now rotate the arch $A^{(0)}$ into the plane  $\pie$,
rotating about its base $v'_0 v'_{i-1}$,
away from $v'_{i-1} v'_i$.  This guarantees that  
$A^{(1)} = A^{(0)} \cup \triangle v'_0 v'_{i-1} v'_i$ is a planar 
weakly-simple polygon.
Moreover, while $\triangle v'_0 v'_{i-1} v'_i$ 
may be degenerate, 
the chain
$(v'_0 ,\ldots, v'_i)$ lies strictly to one side of the line
through $(v'_0,v'_{i-1})$ and so is simple.
See Fig.~\figref{A1}.
\begin{figure}[htbp]
\begin{center}
\ \psfig{figure=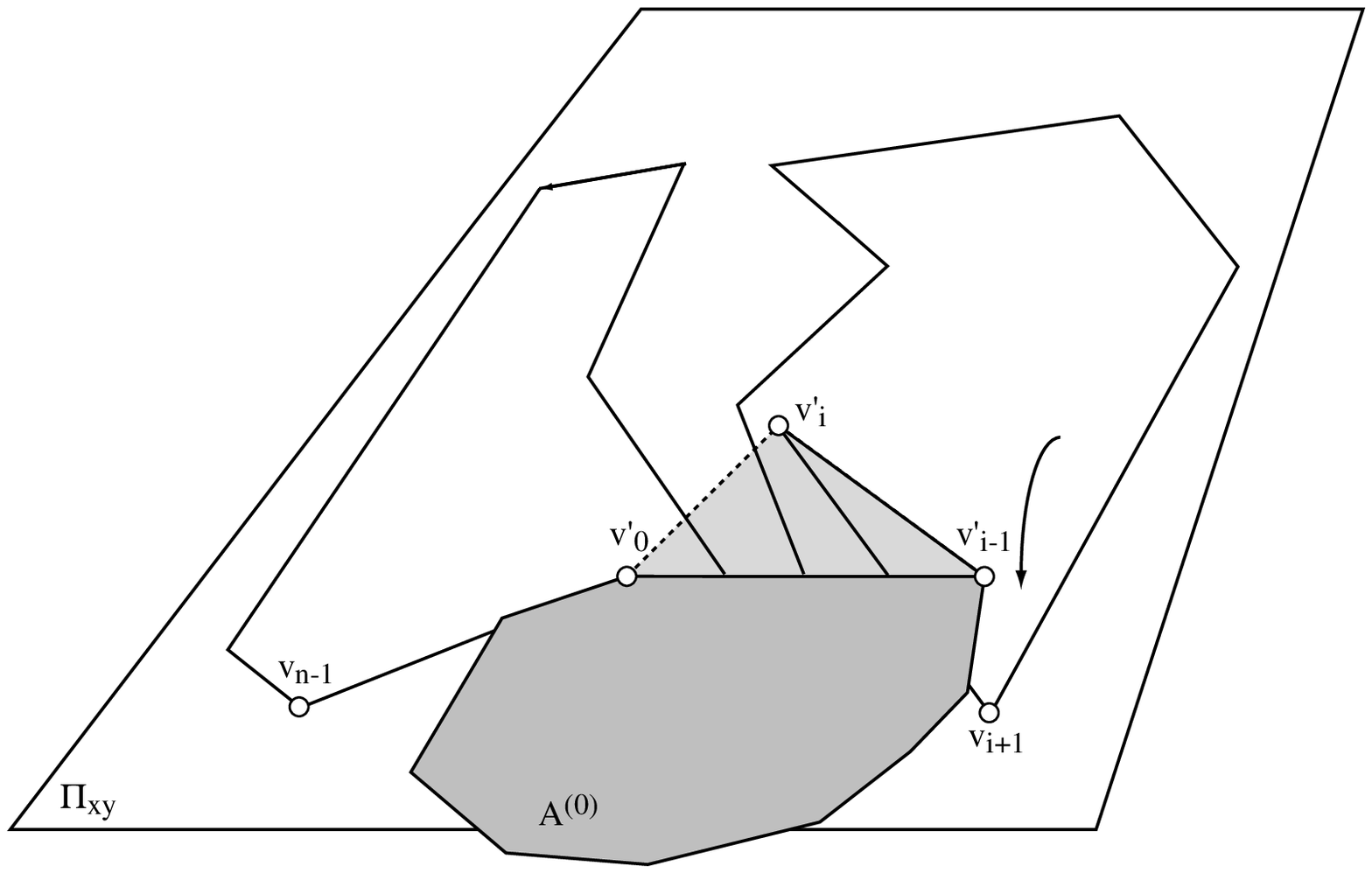,width=10cm}
\conf{
\ \psfig{figure=A1.eps,width=8cm}
}
\end{center}
\caption{$A^{(1)} = A^{(0)} \cup \triangle v'_0 v'_{i-1} v'_i$ lies
in the plane $\pie$ just slightly above $\pixy$.
}
\figlab{A1}
\end{figure}

\subsection{S3}
Now that $A^{(1)}$ lies in its ``own'' plane $\pie$,
it may be convexified without worry about intersections
with the remaining polygon $P[i+1,n-1]$ in $\pixy$.
The polygon $A^{(1)}$ is a ``barbed polygon'': one
that is a union of a convex polygon ($A^{(0)}$) and a triangle
($\triangle v'_0 v'_{i-1} v'_i$).
We establish in Theorem~\theoref{barb.strict}
that $A^{(1)}$ may be convexified in such a way
that neither $v'_0$ nor $v'_i$ move, and 
$v'_0$ and $v'_i$ end up strictly convex vertices of 
the resulting convex polygon $A^{(2)}$.

\subsection{S4}
We next rotate $A^{(2)}$ up into the vertical plane $\Pi_z(v'_0,v'_{i})$.
Because of strict convexity at $v'_0$ and $v'_i$, the arch stays
above $\pie$.
See Fig.~\figref{A2}.
\begin{figure}[htbp]
\begin{center}
\ \psfig{figure=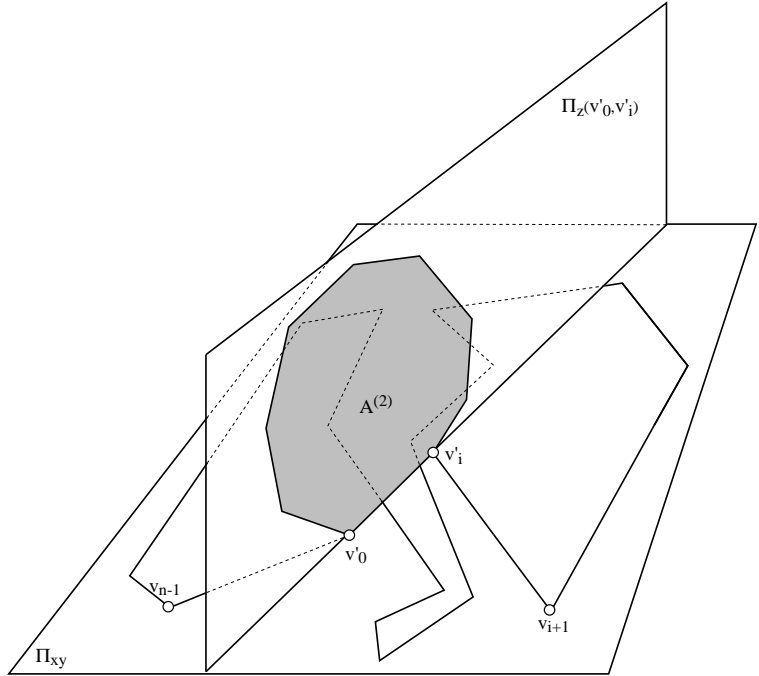,width=10cm}
\conf{
\ \psfig{figure=A2.eps,width=8cm}
}
\end{center}
\caption{$A^{(2)}$, which has incorporated the
edge $v_{i-1} v'_i$ of $P$, is rotated up into the plane $\Pi_z(v'_0,v'_{i})$.
}
\figlab{A2}
\end{figure}

We have now reestablished the induction hypothesis conditions
H1--H5.
After the penultimate step, for $i=n{-}2$, 
only $v_{n-1}$ lies on $\pixy$, and 
the final execution of the lifting Step S1 rotates $v_{n-1}$
about $v'_0 v'_{n-2}$ to raise it to $\pie$.
A final execution of Steps S1 and S2 yields a convex polygon.
Thus, assuming Theorem~\theoref{barb.strict}
in Section~\secref{barbed} below,
we have established the correctness of the algorithm:

\begin{theorem}
The ``St.~Louis Arch'' Algorithm convexifies a planar simple polygon
of $n$ vertices.
\theolab{StLouis}
\end{theorem}

We will analyze its complexity in Section~\secref{Complexity.StLouis}.

We now return to Step S3, convexifying a barbed polygon.
We perform the convexification entirely within the plane $\pie$.
We found two strategies for this task.
One maintains $A$ as a convex quadrilateral,
and the goal of Step S3 can be achieved by convexifying the
(nonconvex) pentagon $A^{(1)}$, and then reducing it to a
convex quadrilateral.
Although this approach is possible, we found it somewhat easier
to leave $A$ as a convex $(i{+}1)$-gon,
and prove that
$A^{(1)} = A^{(0)} \cup \triangle v'_0 v'_{i-1} v'_i$ can be convexified.
This is the strategy we pursue in the next two sections.
Section~\secref{quad} concentrates on the base case,
convexifying a quadrilateral, and Section~\secref{barbed}
achieves Theorem~\theoref{barb.strict}, the final piece needed to
complete Step S3.

\subsection{Convexifying Quadrilaterals}
\seclab{quad}
It will come as no surprise that every planar, nonconvex quadrilateral
can be convexified.  Indeed, recent
work has shown that any star-shaped polygon may be
convexified~\cite{elrsw-cssp-98},
and this implies the result for quadrilaterals.
However, because we need several variations on basic quadrilateral
convexification, we choose to develop our results independently,
although relegating some details to the Appendix.

Let $Q = (v_0,v_1,v_2,v_3)$ be a weakly simple, nonconvex quadrilateral,
with $v_2$ the reflex vertex.
By {\em weakly simple\/} we mean that either $Q$ is simple,
or $v_2$ lies in the relative interior of one of the edges
incident to $v_0$ 
(i.e., no two of $Q$'s edges properly cross).
This latter violation of simplicity is permitted so that we can
handle a collapsed triangle inherited
from step S1 of the arch algorithm 
(Section~\secref{Collinearity}).

As before, let $\a_i$ be the smaller of the two (interior and exterior)
angles at $v_i$.
Call a joint $v_i$ {\em straightened\/} if $\a_i = \pi$,
and {\em collapsed\/} if $\a_i = 0$.
All motions throughout this (\secref{quad}) and the next
section (\secref{barbed}) are in 2D.

We will convexify $Q$ with
one motion $M$, whose intuition is as follows;
see Fig.~\figref{quad.0}.
Think of the two links adjacent to the
reflex vertex $v_2$ as constituting a rope.
$M$ then opens the joint at $v_0$ until the rope becomes taut.
Because the rope is shorter than the sum of the lengths of the other
two links, it becomes taut prior to any other ``event.''

Any motion $M$ that transforms a shape such as $Q$ can take on rather
different appearances when different parts of $Q$ are fixed in
the plane, providing different frames of reference for the motion.  
Although all such fixings represent the same intrinsic
shape transformation $M$, when convenient we distinguish two fixings:
$M_{02}$, which fixes the line $L$ containing $v_0 v_2$,
and $M_{03}$, which fixes the line containing $v_0 v_3$.

The convexification motion $M$ is easiest to see when
viewed as motion 
$M_{02}$. Here the two $2$-link chain
$(v_0, v_1, v_2)$ and $(v_0, v_3, v_2)$ perform a 
{\em line-tracking\/} motion~\cite{lw-rltm-92}:
fix $v_0$, and move $v_2$ away from $v_0$ 
along the fixed directed line $L$ containing
$v_0 v_2$, until $v_2$ straightens.

\begin{lemma}
A weakly simple quadrilateral $Q$ nonconvex at $v_2$ may be convexified
by motion $M_{02}$, which 
straightens the reflex joint $v_2$, thereby converting $Q$
to a triangle $T$.
Throughout the motion, all four angles $\a_i$ increase only,
and remain within $(0,\pi)$ until $\a_2=\pi$.
{\em See Fig.~\figref{quad.0}a.}
\lemlab{quad.M02}
\end{lemma}

Although this lemma is intuitively obvious, and implicit
in work on linkages (e.g., \cite{gn-ogp4bm-86}),
we have not found an explicit statement of it in the literature,
and we therefore present a proof in the Appendix
(Lemma~\lemref{linetrack-theorem}).

\begin{figure}[htbp]
\begin{center}
\ \psfig{figure=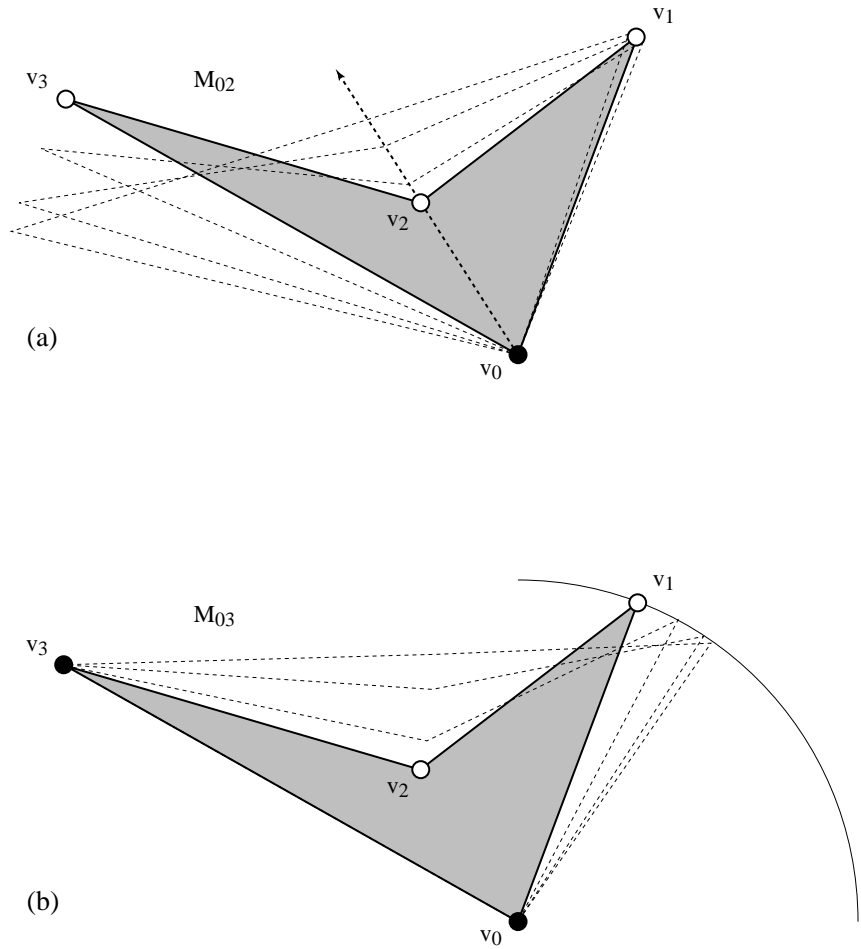,height=5in}
\end{center}
\caption{(a) Convexifying a quadrilateral by $M_{02}$: moving $v_2$ out the
$v_0 v_2$ diagonal;
(b) The same motion viewed as $M_{03}$:
opening $\a_0$ with $v_0 v_3$ fixed.
}
\figlab{quad.0}
\end{figure}
We note that the same motion convexifies a degenerate quadrilateral,
where the triangle $\triangle v_0 v_1 v_2$ has zero area with
$v_2$ lying on the edge $v_0 v_1$.
See Fig.~\figref{quad.degen}.
As long as we open $\a_2$ in the direction, as illustrated,
that makes the quadrilateral simple, the proof of Lemma~\lemref{quad.M02}
carries through.
\begin{figure}[htbp]
\begin{center}
\ \psfig{figure=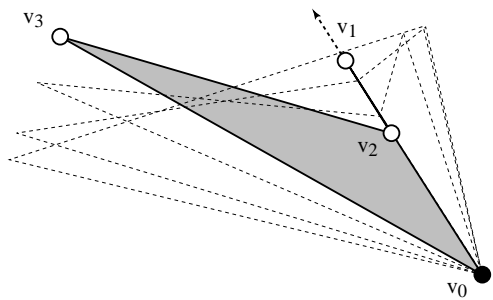,height=4cm}
\end{center}
\caption{Motion $M_{02}$ also convexifies a weakly simple quadrilateral.}
\figlab{quad.degen}
\end{figure}

The motion $M_{02}$ used in Lemma~\lemref{quad.M02} is equivalent to
the motion $M_{03}$ obtained by
fixing $v_0 v_3$ and opening $\a_0$ by
rotating $v_1$ clockwise (cw) around the circle
of radius $\ell_0$ centered on $v_0$.
Throughout this motion, the polygon stays right of the fixed edge $v_0 v_3$.
See Fig.~\figref{quad.0}b.
This yields the following
easy corollary of 
Lemma~\lemref{quad.M02}:
\begin{lemma}
Let $P = Q \cup P'$ be a polygon
obtained by gluing 
edge $v_0 v_3$ of a weakly simple quadrilateral $Q$ nonconvex at $v_2$,
to an equal-length edge of a convex polygon $P'$, such that
$Q$ and $P'$ are on opposite sides of the diagonal $v_0 v_3$.
Then applying the motion $M_{03}$ to $Q$ while keeping $P'$ fixed,
maintains simplicity of $P$ throughout.
\lemlab{quad.halfplane}
\end{lemma}

\subsubsection{Strict Convexity}
Motion $M$ converts a nonconvex quadrilateral into a triangle,
but we will need to convert it to a strictly convex
quadrilateral.  This can always be achieved by continuing $M_{02}$
beyond the straightening of $\a_2$.

\begin{lemma}
Let $Q = (v_0,v_1,v_2,v_3)$ be a quadrilateral, 
with $(v_1,v_2,v_3)$
collinear so that $\a_2=\pi$, and such that $\triangle v_0 v_1 v_3$
is nondegenerate.
As in Lemma~\lemref{quad.halfplane},
let $P = Q \cup P'$ be a convex polygon obtained by gluing $P'$
to edge $v_0 v_3$ of $Q$, with $v_0$ and $v_3$
strictly convex vertices of $P$.
The motion $M_{02}$ (moving $v_2$ along the line determined by $v_0 v_2$)
transforms $Q$ to a strictly convex quadrilateral $Q'$
such that $Q' \cup P'$ remains a convex polygon
{\em (See Fig.~\figref{quad.strict}.)}
\lemlab{quad.strict}
\end{lemma}
\begin{figure}[htbp]
\begin{center}
\ \psfig{figure=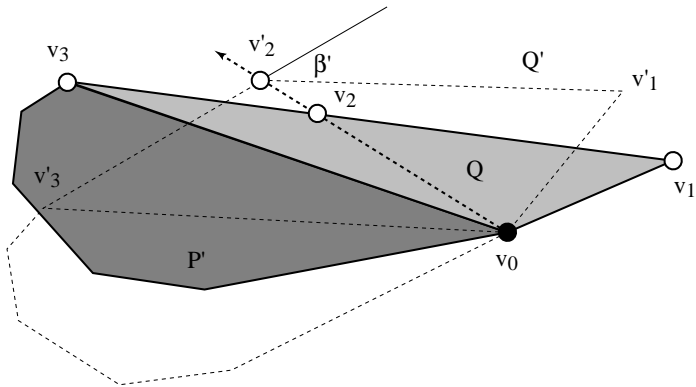,height=2in}
\end{center}
\caption{Converting $Q$ to the strictly convex quadrilateral $Q'$
via $M_{02}$. 
Attachment $P'$ is carried along rigidly.
}
\figlab{quad.strict}
\end{figure}
\begin{pf}
Because $v_0$ and $v_3$ are strictly convex vertices, and $v_1$
must be strictly convex because $Q$ is a nondegenerate triangle, 
all the interior angles at these
vertices are bounded away from $\pi$.  By assumption, they
are also bounded away from $0$.  Thus there is some freedom
of motion for $v_2$ along the line determined by $v_0 v_2$
before the next event, when one of these angles reaches $0$ or $\pi$.
\end{pf}

A lower bound on $\b' = \pi - \a'_2$, the amount that $v_2$
can be bent before an event is reached, could be computed
explicitly in $O(1)$ time
from the local geometry of $Q \cup P'$, but we will not
do so here.

\subsection{Convexifying Barbed Polygons}
\seclab{barbed}

Call a polygon {\em barbed\/} if removal of one ear $\triangle abc$
leaves a convex polygon $P'$.
$\triangle abc$ is called the {\em barb\/} of $P$.
Note that either or both of vertices $a$ and $c$ may be reflex vertices of $P$.
In order to permit $\triangle abc$ to be degenerate (of zero area),
we extend the definition as follows.
A weakly simple polygon
(Section~\secref{quad}, Figure~\figref{quad.degen})
is {\em barbed\/} if, for three consecutive vertices $a$, $b$, $c$,
deletion of $b$ (i.e., removal of the possibly degenerate $\triangle abc$)
leaves a simple convex polygon $P'$.
Note this definition only permits weak simplicity at the barb $\triangle abc$.

The following lemma (for simple barbed polygons) is
implicit in~\cite{s-scsc-73}, 
and explicit (for star-shaped polygons, which includes barbed
polygons) in~\cite{elrss-cssp-98},
but we will need to subsequently
extend it, so we provide our own proof.

\begin{lemma}
A weakly simple barbed polygon may be convexified, with $O(n)$ moves.
\lemlab{barb}
\end{lemma}
\begin{pf}
Let $P = (v_0, v_1, \ldots, v_{n-1})$, with
$\triangle v_0 v_{n-2} v_{n-1}$ the barb.
See Fig.~\figref{barb}.
\begin{figure}[htbp]
\begin{center}
\ \psfig{figure=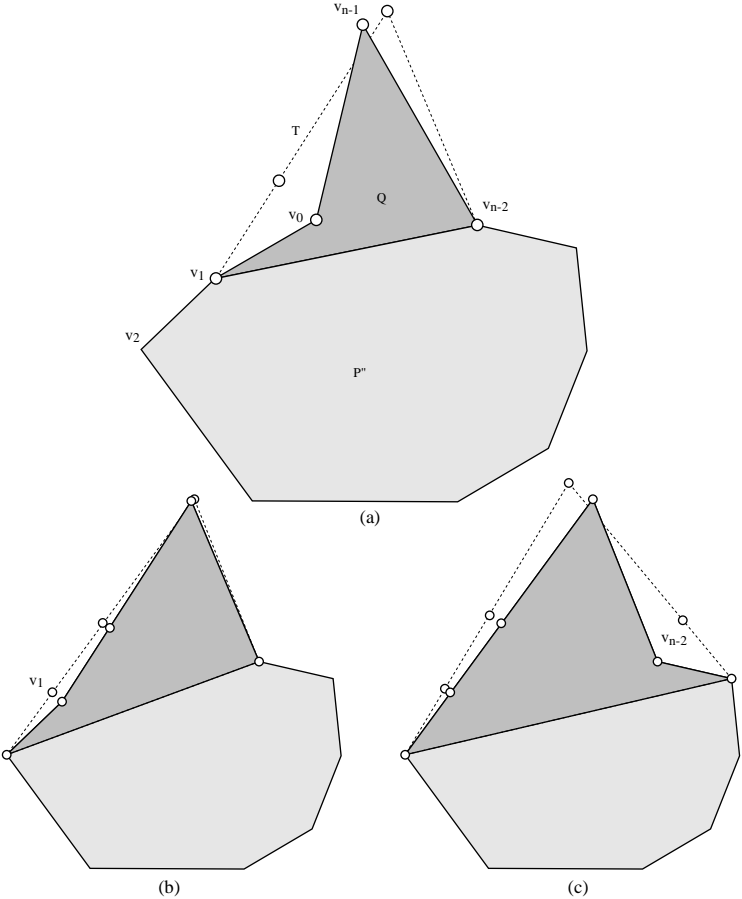,height=12cm}
\end{center}
\caption{(a) A barbed polygon with barb $\triangle v_0 v_{n-2} v_{n-1}$.
The nonconvex quadrilateral $Q$ is transformed to $T$, resulting
in a new barbed polygon $T \cup P''$.
(b) and (c) show the remaining convexification steps.}
\figlab{barb}
\end{figure}

The proof is by induction.
Lemma~\lemref{quad.M02} establishes the base case, $n=4$,
for every quadrilateral is a barbed polygon.
So assume the theorem holds for all barbed polygons of up to $n-1$
vertices.

If both $v_0$ and $v_{n-2}$ are convex, $P$ is already convex and we
are finished.
So assume that $P$ is nonconvex, and without loss of generality
let $v_0$ be reflex in $P$.
It must be that $v_1 v_{n-2}$ is a diagonal, as it lies within
the convex portion of $P$.
Let $Q = (v_0, v_1, v_{n-2}, v_{n-1})$ be the quadrilateral cut off
by diagonal $v_1 v_{n-2}$,
and let $P'' = (v_1,\ldots,v_{n-2})$ be the remaining portion of
$P$, so that $P = Q \cup P''$.
$Q$ is nonconvex at $v_0$.

Lemma~\lemref{quad.halfplane} shows that motion $M$ (appropriately
relabeled) may be applied to convert $Q$ 
to a triangle $T$ by straightening $v_0$,
leaving $P''$ unaffected.
At the end of this motion, we have reduced $P$ to a polygon $P'$
of one fewer vertex.
Now note that $T$ is a barb for $P'$ (because $P''$ is convex):
$P' = T \cup P''$.
Apply the induction hypothesis to $P'$.
The result is a convexification of $P$.

Each reduction uses one move $M$, and so $O(n)$ moves suffice for $P$.
\end{pf}

Note that although each step of the convexification straightens
one reflex vertex, it may also introduce a reflexivity:
$v_1$ is convex in Fig.~\figref{barb}a but reflex in 
Fig.~\figref{barb}b.  We could make the procedure more
efficient by ``freezing'' any joint as soon as it straightens,
but it suffices for our analysis to freeze each straightened
reflex vertex, thenceforth treating the segment on which it lies
as a single rigid link.

As is evident in 
Fig.~\figref{barb}c, the convexification leaves a polygon 
with several vertices straightened.  One of the edges $e$ of the
barbed polygon is the base of the arch $A$ from Section~\secref{S1.lifting}.
If either of $e$'s endpoints are straightened, then part
of the arch will lie directly in the plane $\pie$, and could
cause a simplicity violation during the S1 lifting step.
Therefore we must ensure that both of $e$'s endpoints are
strictly convex:

\begin{lemma}
Any convex polygon with a distinguished edge $e$ can be
reconfigured so that
that both endpoints of $e$
become strictly convex vertices.
\lemlab{strict}
\end{lemma}
\begin{pf}
Suppose the counterclockwise endpoint $v_2$ of $e$ 
has internal angle $\a=\pi$;
see Fig.~\figref{barb.strict}.
Let $v_1$ be the next strictly convex vertex in clockwise order before $v_2$
(it may be that $v_1$ is the other endpoint of $e$),
and $v_3, v_0$ be the next two strictly 
convex vertices adjacent to
$v_2$ counterclockwise.
Let $Q=(v_0,v_1,v_2,v_3)$. 
\begin{figure}[htbp]
\begin{center}
\ \psfig{figure=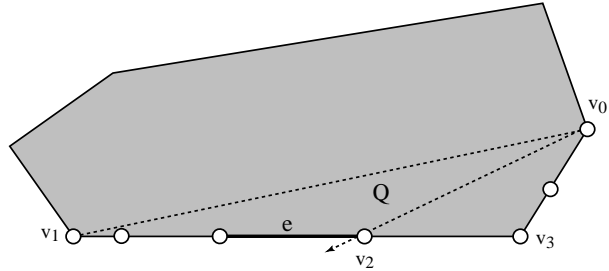,width=8cm}
\end{center}
\caption{Making one endpoint of $e$ strictly convex.}
\figlab{barb.strict}
\end{figure}
Then apply Lemma~\lemref{quad.strict} to $Q$ to convexify $v_2$
via motion $M_{02}$.
Apply the same procedure to the other endpoint of $e$
if necessary.
\end{pf}

Using Lemma~\lemref{barb} to convexify the barbed polygon arch,
and Lemma~\lemref{strict} to make its base endpoints strictly
convex, yields:
\begin{theorem}
A weakly simple barbed polygon may be convexified in such a
way that the endpoints of a distinguished edge are strictly convex.
\theolab{barb.strict}
\end{theorem}

This completes the description of the St.\ Louis Arch Algorithm,
as $A^{(1)} = A^{(0)} \cup \triangle v'_0 v'_{i-1} v'_i$ is a barbed
polygon, and Step S4 may proceed because of the strict convexity
at the arch base endpoints.

\subsection{Complexity of St.\ Louis Arch Algorithm}
\seclab{Complexity.StLouis}
It is not difficult to see that only a constant number of moves
are used in steps S0, S1, S2, and S4.
Step S3 is the only exception, which we have seen
in Lemma~\lemref{barb}
can be executed in $O(n)$ moves.
So the resulting procedure can be accomplished in $O(n^2)$ moves. 
The algorithm actually only uses $O(n)$ moves, as the following
amortization argument shows:
\begin{lemma}
The St.\ Louis Arch Algorithm runs in $O(n)$ time and
uses $O(n)$ moves.
\end{lemma}
\lemlab{amort}
\begin{pf}
Each barb convexification move used in the proof of Lemma~\lemref{barb}
constitutes a single move according to the definition in
Section~\secref{Complexity}, as four joints open monotonically
(cf.~Lemma~\lemref{quad.M02}).
Each such convexification move necessarily straightens one reflex
joint, which is subsequently ``frozen.''
The number of such freezings is at most $n$ over the life of
the algorithm.  So although any one barbed polygon might
require $\Omega(n)$ moves to convexify, the convexifications
over all $n$ steps of the algorithm uses only $O(n)$ moves.
Making the base endpoint angles strictly convex requires
at most two moves per step, again $O(n)$ overall.

Each step of the algorithm can be executed in constant time,
leading to a time complexity of $O(n)$.
Again we must consider computation of the minimum distances
around each vertex to obtain $\d$ (Section~\secref{delta}),
but we can employ the same medial axis technique used
in Section~\secref{Open.3D} to compute these distances
in $O(n)$ time.
\end{pf}

Note that at most four joints rotate at any one time,
in the barb convexification step.

\section{Open problems}
\seclab{Open}
Although we have mapped out some basic distinctions between
locked and unlocked chains in three dimensions, our results leave many
aspects unresolved:
\begin{enumerate}
\conf{\squeezelist}
\item What is the complexity of deciding whether a chain in 3D can be unfolded?

\item Theorem~\theoref{simple.proj} only covers
chains with simple orthogonal projections.
Extension to 
perspective (central) projections, or 
other types of projection,
seems possible.

\item
Can a closed chain with a simple projection always be convexified?
None of the algorithms presented in this paper seem to settle
this case.

\item 
Find unfolding algorithms that minimize the number
of simultaneous
joint rotations.
Our quadrilateral convexification procedure,
for example, moves four joints at once, whereas pocket flipping
moves only two at once.

\item Can an open chain of unit-length links lock in 3D?
Cantarella and Johnson show in~\cite{cj-nepiu-98}
that the answer is {\sc no} if $n \le 5$.

\end{enumerate}

\subsection*{Acknowledgements}
We thank 
W.~Lenhart for co-suggesting the knitting needles
example in Fig.~\figref{knitting},
J.~Erickson for the amortization argument that reduced the time
complexity in Lemma~\lemref{amort} to $O(n)$,
and H.~Everett for useful comments.

\appendix
\section{Appendix}
\subsection{Computation of $\epsilon$}
Here we detail a possible computation of $\epsilon$,
as needed in 
Section~\secref{Epsilon}.

The smallest radius $r$ for the circle $C$ is determined by the
minimum angle $\b$
(the smallest deviation from straightness)
and the shortest edge length $\ell$.
In particular, $r \ge \ell \sin(\b/2)$; see
Figure~\figref{trigeom}a,b.
Here it is safe to use the $\b$ from the plane $\pixy$
because the deviation from straightness is only larger
in the tilted plane of $\triangle v_{i+1},v_i,v'_{i-1}$
(cf.~Fig.~\figref{double.cone}),
and we seek a lower bound on $r$.
\begin{figure}[htbp]
\begin{center}
\ \psfig{figure=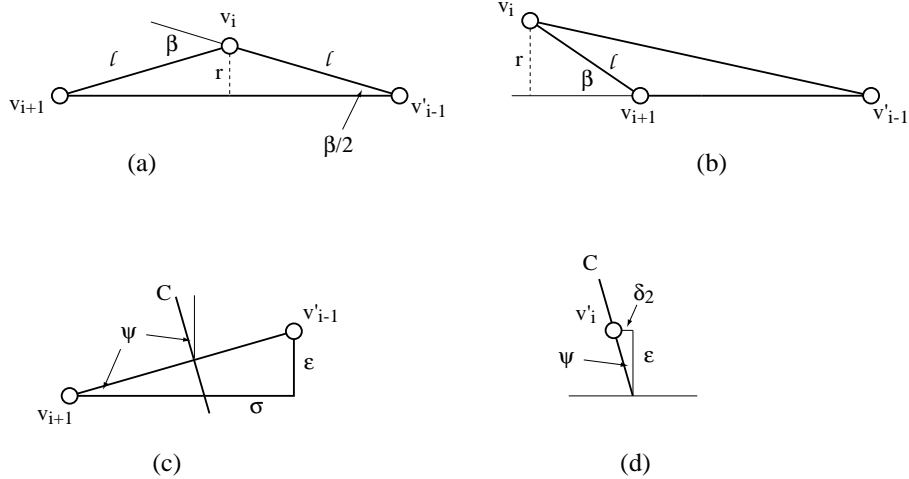,width=12cm}
\end{center}
\caption{
Determination of smallest circle radius $r$:
(a) $r \ge \ell \sin(\b/2)$; (b) $r \ge \ell \sin\b$.
So $\ell \sin(\b/2)$ is a lower bound.
Determination of largest circle tilt $\psi$:
(c) $\cos \psi \ge \s / \protect\sqrt{\s^2+\e^2}$.
(d) Determination of $\d_2$: $\d_2 \le \e \tan \psi$.
}
\figlab{trigeom}
\end{figure}

The tilt $\psi$ of the circle leaves the top of $C$ at least at height 
$r \cos \psi$.  
Because $|v'_{i-1} v_{i+1}| \ge \s$, the tilt angle must satisfy
$\cos \psi \ge \s / \sqrt{\s^2+\e^2}$;
see Figure~\figref{trigeom}c.
Thus to meet condition~(1), we should arrange that
$$
\ell \sin(\b/2) {\frac{\s}{\sqrt{\s^2+\e^2}}} > \e
$$
which can clearly be achieved by choosing $\e$ small enough,
as $\ell$, $\b$, and $\s$ are all constants fixed by the geometry of $P$.

Turning to condition~(2) of Section~\secref{Epsilon}, 
the movement of $v'_i$ with respect to $v_i$
can be decomposed into two components.
The first is determined by the rotation along $C$ if that circle were
vertical.
This deviation is no more than $\d_1 = r (1 - \cos \phi)$,
where $\phi$ is the lifting rotation angle measured at the cone axis
$v'_{i-1} v_{i+1}$.
Because $\sin \phi \le \e/r$, this leads to
$\d_1 \le r \left[ 1 - \sqrt{ 1 - (\e / r)^2 } \right]$.
The second component is due to the tilt of the circle,
which is 
$\d_2 = \e \tan \psi \le \e^2 / \s$; see Figure~\figref{trigeom}d.
The total displacement is no more than $\d_1+\d_2$.
Now it is clear that as $\e \rightarrow 0$, both 
$\d_1 \rightarrow 0$ and
$\d_2 \rightarrow 0$.
Thus for any given $\d$, we may choose $\e$ such that $\d_1+\d_2 < \d$.

\subsection{Straightening Lemma}

The following lemma is used to determine $\d$ in Section~\secref{delta}.
\begin{lemma}
Let $ABC$ be a triangle, with $|AB| \ge \ell$, $|BC| \ge \ell$,
and $\b \le \angle ABC \le \pi - \b$.
Then for any triangle $A'B'C'$ whose vertices are displaced
at most $\d$ from those of $\triangle ABC$,
i.e.,
$$
|AA'| \: < \: \d, \;\;
|BB'| \: < \: \d, \;\;
|CC'| \: < \: \d, 
$$
$\angle A'B'C' < \pi$.
\lemlab{rho}
\end{lemma}
\begin{pf}
Let 
$a$ be the point on $BA$ a distance $\ell/2$ from $B$,
and let
$c$ be the point on $BC$ a distance $\ell/2$ from $B$.
Let $L$ be the line containing $ac$.
Set $\t = \angle Bac = \angle acB$, and $\phi = \angle aBc = \pi - 2 \t$.
Because $\phi = \angle ABC$, the assumptions of the lemma
give 
$\b \le \pi - 2 \t \le \pi - \b$, or 
$\b/2 \le \t \le (\pi-\b)/2$.
The distance $d(A,L)$ from $A$ to $L$ satisfies
\begin{eqnarray}
d(A,L) & \ge & (\ell/2) \sin \t \\
       & \ge & (\ell/2) \sin (\b/2) \\
       & >   & \d \;.
\end{eqnarray}
The exact same inequality hold for the distances $d(B,L)$
and $d(C,L)$, because the relevant angle is $\t$ is each case,
and the relevant hypotenuse is $\ge \ell/2$ in each case.
\begin{figure}[htbp]
\begin{center}
\ \psfig{figure=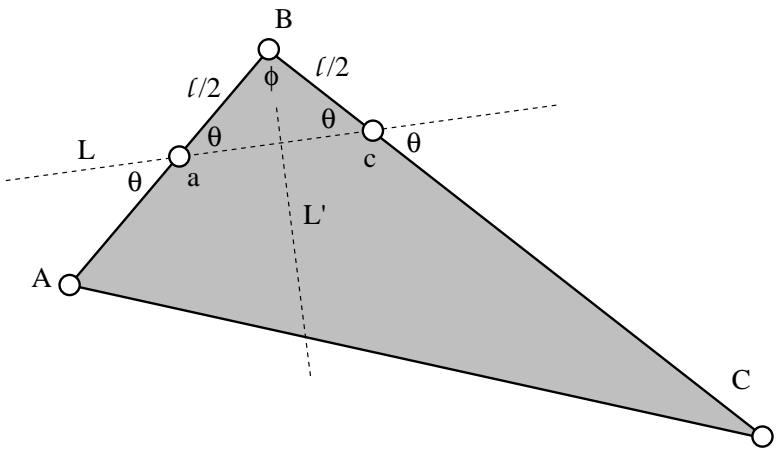,height=6cm}
\end{center}
\caption{$A$ and $C$ are separated by $L$ from $B$, 
and by $L'$ from each other.}
\figlab{delta}
\end{figure}

Now suppose the three vertices 
$A'$, $B'$, $C'$ each move no more
than $\d$ from 
$A$, $B$, and $C$ respectively.
Then $L$ continues to separate $A'$ and $C'$ from $B'$,
by the above argument.

%
\end{pf}

\subsection{Quadrilateral Convexification}
The next results are employed in Section~\secref{quad}
on convexifying quadrilaterals.
We need the following lemma that states that the
reflex joint of a quadrilateral
can be straightened in the first place.
Let $Q=v_0v_1v_2v_3$ be a four-bar linkage with $v_2$ a reflex
joint.

\begin{lemma} \label{four-bar-linkage-lemma}
A non-convex four-bar linkage can be convexified into a
triangle by straightening its reflex joint.
\end{lemma}
\begin{pf}
Let $ray(v_0,v_2)$ be the ray starting at $v_0$ in the direction of $v_2$ and refer
to Figure~\figref{linetrack}.
Without loss of generality let $v_0$ be the origin, $ray(v_0,v_2)$
the positive $x$-axis 
and assume the sum of the link lengths $(l_0+l_1)$ is smaller than $(l_2+l_3)$.
Assume that $v_2$ is translated continuously along the $x$-axis in the positive
direction until it gets stuck. Since $v_2$ cannot move further
it follows that joints $v_0$, $v_1$ and $v_2$ all lie on the
$x$-axis and joint $v_1$ has been straightened.
This implies the new interior angle of $v_2$, $\gamma < \pi$.
But before the motion $v_2$ was a reflex angle with $\gamma > \pi$.
Since the angles change continuously there must exist a point
during the motion at which $\gamma = \pi$.
\end{pf}

\begin{lemma} 
When $d(v_0,v_2)$ is increased, all joints of the linkage open,
that is, the interior angles of the convex joints and the
exterior angle of the reflex joint all increase. 
\lemlab{linetrack-theorem}
\end{lemma}
\begin{pf}
We will show that if $v_2$ is moved along $ray(v_0, v_2)$ in such a way
that the distance $d(v_0, v_2)$ is increased by some positive real
number $\epsilon$, no matter how small, while $v_2$ remains reflex,
then all joints open.
First note that by Euclid's Proposition 24 of Book I,
$v_1$ and $v_3$ open, that is, their interior angles increase.
Secondly, note that if the interior angle at $v_0$ opens
then so does the exterior angle at $v_2$ (and vice-versa)
by applying Euclid's Proposition 24 to distance $d(v_1, v_3)$.
Hence another way to state the theorem 
in terms of distances only is: in a non-convex
four-bar linkage the length of the interior diagonal increases if,
and only if, the length of the exterior diagonal increases.
It remains to show that
increasing $d(v_0,v_2)$ increases the angle at $v_0$.

Before proceeding let us take care of the value of $\epsilon$.
While there is no problem selecting $\epsilon$ too small,
we must ensure it is not too big, for otherwise
when we increase $d(v_0, v_2)$ by $\epsilon$
the linkage may become convex. From Lemma~\ref{four-bar-linkage-lemma}
we know that as $d(v_0, v_2)$ is increased the linkage will become
a triangle at some point when joint $v_2$ straightens,
at which time $d(v_0, v_2)$ will have reached its maximum value, say $l$.
Using the law of cosines for this triangle we obtain 

\[l^2 = l^2_2 + l^2_3 - l_2\{[(l_1 + l_2)^2 + l^2_3 - l^2_0]/(l_1 + l_2)\}.\]

\noindent Therefore if we choose $\epsilon$ such that

\[\epsilon < l - d(v_0, v_2),\]

\noindent then we ensure that $v_2$ remains reflex.

It is convenient to analyse the situation with link $v_3v_0$ as
a rigid frame of reference rather than the $ray(v_0, v_2)$.
Therefore let both $v_0$ and $v_3$ be fixed in the plane.
Then as $d(v_0, v_2)$ increases, from Euclid's Proposition 24
it follows that $v_1$ rotates about $v_0$ along the {\em fixed\/}
circle $C(v_0,l_0)$ centered at $v_0$ of radius $l_0$,
$v_2$ rotates about $v_3$ on the {\em fixed\/} circle $C(v_3,l_2)$ centered
at $v_3$ with radius $l_2$, and  $ray(v_0, v_2)$ rotates about
$v_0$.

Denote the initial configuration by $Q=v_0v_1v_2v_3$
and the final configuration after $d(v_0, v_2)$ is increased
by $\epsilon$ by $Q' = v_0u_1u_2v_3$.
In other words $v_1$ has moved to $u_1$, $v_2$ has moved to
$u_2$ and $ray(v_0, v_2)$ has moved to $ray(v_0, u_2)$.
Since the exterior angle at $v_2$ is less than $\pi$
and link $v_3v_2$ rotates in a counterclockwise manner
this motion causes $u_2$ to penetrate the interior
of the shaded circle $C(v_1, l_1)$ centered at $v_1$
with radius $l_1$. 
Furthermore, $u_2$ cannot overshoot this shaded circle and find
itself in its exterior after having penetrated it, for this
would imply the joint $u_2$ is convex, which is impossible
for the value of $\epsilon$ we have chosen.
Now, since $u_2$ is in the interior of the shaded disk
$C(v_1, l_1)$ and the radius of this disk is $l_1$
it follows that the distance $d(u_2, v_1)$
is less than the link length $l_1$.
Let us therefore extend the segment $u_2v_1$ along the 
$ray(u_2,v_1)$ to a point $u'_2$ so that $d(u_2, u'_2)
= l_1$. 
	Note that the figure shows the situation
        when $u_2'$ lies in the exterior of $C(v_0,l_0)$.
        If $u_2'$ lies on $C(v_0,l_0)$ it yields $u_1$
        imediately. If $u_2'$ lies in the interior
        of $C(v_0,l_0)$ then the arc $u_1,u_2',u_1'$
        in the figure would be in the interior of
        $C(v_0,l_0)$.
But of course $u_1$, the new position of $v_1$, must
lie on the circle $C(v_0, l_0)$.
To compute the possible locations for  $u_1$ we rotate
segment $u_2u'_2$ about $u_2$ in both the clockwise and
counterclockwise directions to intersect
the circle  $C(v_0, l_0)$ at points $u_1$ and $u'_1$, respectively.
Since $u_2$ lies on $ray(v_0, u_2)$ it follows that $u'_1$
lies to the left of $ray(v_0, u_2)$.
But the two links $v_0u_1$ and $u_1u_2$ must remain
to the right of $ray(v_0, u_2)$ because the links are not
allowed to cross each other. Therefore $u'_1$ cannot be the final position
of link $v_1$ and the latter must move to $u_1$.
Now since 

\[ d(u_2, u'_2) = l_1 > d(u_2, v_1),\]

\noindent it follows that $u_1$ lies clockwise from $v_1$.
Therefore link $v_0v_1$ has rotated clockwise with
respect to $v_0$ and since link $v_0v_3$ is fixed
the interior angle at $v_0$ has increased, proving the theorem.
\end{pf}

\begin{figure}[htbp]
\begin{center}
\ \psfig{figure=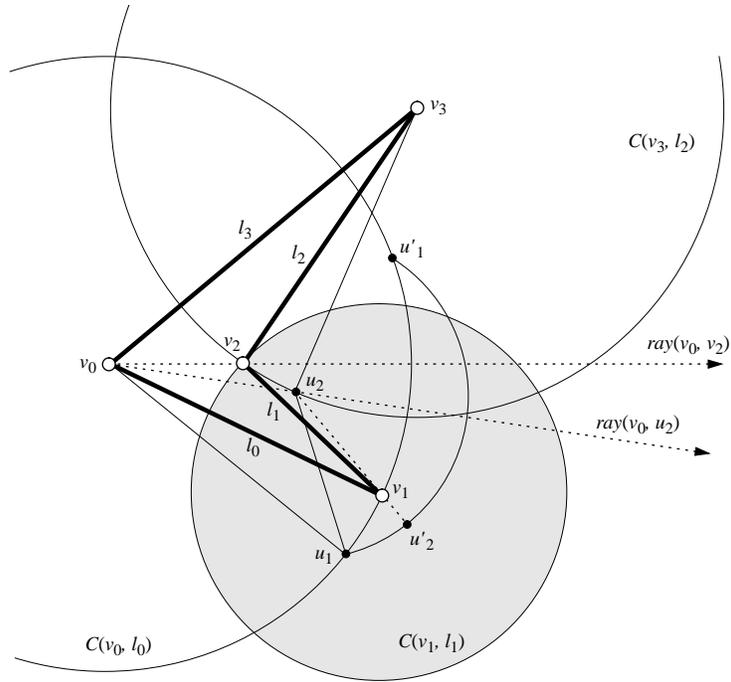,height=15cm}
\end{center}
\caption{Line tracking holding link $v_0v_3$ fixed.
}
\figlab{linetrack}
\end{figure}


\bibliographystyle{alpha}
\bibliography{locked,/home1/orourke/bib/geom/geom}

\conf{
\vspace*{1cm}
\small \noindent
{\bf Acknowledgements}.
We thank 
W.~Lenhart for co-suggesting the knitting needles
example in Fig.~\figref{knitting},
and H.~Everett for useful comments.
This research was initiated at 
a workshop
at the Bellairs Res. Inst. of McGill Univ., 
Jan. 31--Feb. 6, 1998.
Reseach supported in part by FCAR, NSERC, and NSF.
\normalsize
}

\end{document}